\documentclass[acmsmall]{acmart}

\usepackage{array,multirow,etoolbox,graphicx,subcaption,fancyhdr,tabularx,longtable,enumitem}
\usepackage{pifont}
\usepackage{csquotes}
\usepackage{caption}
\usepackage{textcomp}
\usepackage{siunitx}
\usepackage{float}
\usepackage{xspace}
\usepackage{tabularx}
\usepackage{svg}

\usepackage{color}
\definecolor{lightgray}{gray}{0.75}

\usepackage{array,booktabs,ragged2e}
\newcolumntype{R}[1]{>{\RaggedLeft\arraybackslash}p{#1}}

\AtBeginDocument{%
  \providecommand\BibTeX{{%
    \normalfont B\kern-0.5em{\scshape i\kern-0.25em b}\kern-0.8em\TeX}}}

\setcopyright{rightsretained}
\ccsdesc[500]{Software and its engineering~Integrated and visual development environments}
\ccsdesc[500]{Software and its engineering~Software maintenance tools}
\ccsdesc[500]{Software and its engineering~Software configuration management and version control systems}

\keywords{Pull-based software development, pull request, merge conflict, distributed software development}

\setcopyright{rightsretained}
\acmJournal{TOSEM}

\begin{document}

\newlist{steps}{enumerate}{1}
\setlist[steps, 1]{label = Step \arabic*:}

\author{Chandra Maddila}
\affiliation
{
\institution{Microsoft Research}
\city{Redmond}
\state{WA, USA}
}
\email{chmaddil@microsoft.com}

\author{Nachiappan Nagappan}
\additionalaffiliation{%
\institution{Microsoft Research}
}
\affiliation
{
\institution{Microsoft Research}
\city{Redmond}
\state{USA}
}
\email{nchiappan.nagappan@gmail.com}

\author{Christian Bird}
\affiliation
{
\institution{Microsoft Research}
\city{Redmond}
\state{WA, USA}
}
\email{cbird@microsoft.com}

\author{Georgios Gousios}
\additionalaffiliation{%
\institution{Delft University of Technology}
\city{Delft}
\state{The Netherlands}
}
\affiliation
{
\institution{Facebook}
\city{Menlo Park}
\state{CA, USA}
}
\email{gousiosg@fb.com}

\author{Arie \lowercase{van} Deursen}
\orcid{0000-0003-4850-3312}
\affiliation
{
\institution{Delft University of Technology}
\city{Delft}
\state{The Netherlands}
}
\email{Arie.vanDeursen@tudelft.nl}


\newcommand{\cone}{ConE\xspace}
\newcommand{\pr}{pull request\xspace}
\newcommand{\csharp}{C\#}
\newcommand{\prl}{PullRequest Lifetime}
\newcommand{\ce}{Concurrent Edit}
\newcommand{\cmark}{\ding{51}}%
\newcommand{\xmark}{\ding{55}}%
\newcommand{\etal}{\emph{et al.}\xspace}
\newcommand{\NumDecorations}{775\xspace}
\newcommand\NumRepos{234}
\newcommand\NumPrs{26,000}
\newcommand\NumDevs{30,000}
\newcommand{\PositiveResponseRate}{71.48\xspace}
\begin{abstract}
Modern, complex software systems are being continuously extended and adjusted.
The developers responsible for this may come from different teams or organizations, and may be distributed over the world. This may make it difficult to keep track of what other developers are doing,
which may result in multiple developers concurrently editing the same code areas.
This, in turn, may lead to hard-to-merge changes or even merge conflicts, logical bugs that are difficult to detect,
duplication of work, and wasted developer productivity.
To address this, we explore the extent of this problem in the pull request based software development model.
We study half a year of changes made to six large repositories in Microsoft in which at least 1,000 pull requests are created each month.
We find that files concurrently edited in different pull requests are more likely to introduce bugs.
Motivated by these findings, 
we design, implement, and deploy 
a service named ConE (Concurrent Edit Detector)
that proactively detects pull requests containing concurrent edits,
to help mitigate the problems caused by them.
ConE has been designed to scale, and to minimize false alarms while still flagging relevant concurrently edited files.
Key concepts of ConE include the detection of the \emph{Extent of Overlap} between pull requests, and the identification of \emph{Rarely Concurrently Edited Files}.
To evaluate ConE, we report on its operational deployment on 234 repositories inside Microsoft.
ConE assessed 26,000 pull requests and made 775 recommendations about conflicting changes, which were rated as useful in over 70\% (554) of the cases.
From interviews with 48 users we learned that they believed \cone would save time in conflict resolution and avoiding duplicate work, and that over 90\% intend to keep using the service on a daily basis.



\end{abstract}

\title{ConE: A Concurrent Edit Detection Tool for Large Scale Software Development}

\maketitle

\section{Introduction} \label{Intro}
In a collaborative software development environment, developers, commonly, work on their individual work items independently by forking a copy of the code base from the latest main branch and editing the source code files locally. They then create \pr{}s to merge their local changes into the main branch. With the rise of globally distributed and large software development teams, this adds a layer of complexity due to the fact that developers working on overlapping parts of the same codebase might be in different teams or geographies or both. While such collaborative software development is essential for building complex software systems that meet the expected quality thresholds and delivery deadlines, it may have unintended consequences or `side effects' \cite{10.1145/2597073.2597074, 5069475, 8094445, 10.1007/s10664-017-9586-1}. The side effects can be as simple as syntactic merge conflicts, which can be handled by version control systems \cite{10.1109/TSE.1975.6312866} and various techniques/tools \cite{BreakingtheCode, IntegratingPrograms, ConfigurationManagementTool}, to semantic conflicts \cite{SemanticConflicts}. Such bugs can be very hard to detect and may cause substantial disruptions \cite{IntegratingPrograms}. Primarily, all of this happens due to lack of awareness and early communication among developers editing the same source code file or area, at the same time, through active \pr{}s.

There is no substitute to resolving merge or semantic conflicts (or fixing logical bugs or refactoring duplicate code) when the issue is manifested. Studies show that \pr{}s getting into merge conflicts is a prevalent problem \cite{10.5555/2486788.2486884, inproceedings-Holmes, articleconflict}. Merge conflicts have a significant impact on code quality and can disrupt the developer workflow \cite{6450947, 10.1145/958160.958177, 8170085}. Sometimes, the conflict becomes so convoluted that one of the developers involved in the conflict has to abandon their change and start afresh. Because of that, developers often defer resolving their conflicts \cite{article-Nelson} which makes the conflict resolution even harder at a later point of time \cite{book-conflict, article-Nelson}. Time spent in conflict resolution or refactoring activities is going to take away valuable time and prohibits developers from fulfilling their primary responsibility, which is to deliver value to the organization in the form of new functionality, bug fixes and maintaining the service. In addition to loss of time and money, this causes frustration \cite{Palantir, 10.1145/2393596.2393648}. Studies have shown that these problems can be avoided by following strategies such as effective communication within the team \cite{CSCW2015GuzziBRD}, and developing awareness about others' changes that have a potential to incur conflicts \cite{6915251}.

Our goal is to design a method to help developers discover changes made on other branches that might conflict with their own changes. This goal is particularly challenging for modern, large scale software development, involving thousands of developers working on a shared code base. One of the design choices that we had to make was to minimize the false alarms by making it more conservative. Studies have shown that, in large organizations, tools that generate many false alarms are not used and eventually deprecated \cite{GoogleLessons}.

The direct source of inspiration for our research is complex, large scale software development as taking place at Microsoft. Microsoft employs \char`\~ 166K employees worldwide and 58.6\% of Microsoft's employees are in engineering organizations. Microsoft employs  \char`\~ 69K employees outside of the United States making it truly multinational \cite{Microsoft-Facts}. Because of the scale and breadth of the organization, tools and technologies used across the company, it is very common for Microsoft's developers to constantly work on overlapping parts of the source code, at the same time, and encounter some of the problems explained above. 

Over a period of twelve months, we studied \pr{}s, source control systems, code review tools, conflict detection processes, and team and organizational structures, across Microsoft and across different geographies. This greatly helped us assess the extent of the problem and practices followed to mitigate the issues induced by the collaborative software development process. We make three key observations:

\begin{enumerate}

\item \textit{Discovering others' changes is not trivial.} There are several solutions offered by source control systems like GitHub or Azure DevOps \cite{GitHub, AzDo} that enable developers to subscribe to email notifications when new \pr{}s are created or existing ones are updated. In addition, products like Microsoft Teams or Slack can show a feed of changes that are happening in a repository a user is interested in. The notification feed becomes noisy over time and it becomes very hard for developers to digest all of this information and locate \pr{}s that might cause conflicts.
This problem is aggravated when a developer works on multiple repositories.

\item \textit{Tools have to fit into developers' workflows.} Making developers install several client tools and making them switch their focus between different tools and windows is a big obstacle for adoption of any solution. There exists a plethora of tools \cite{Palantir, FASTDash, EarlyDetection-2/2025113.2025139} that aim to solve this problem in bits and pieces. Despite this, usability is still a challenge because none of them fit naturally into developers' workflows. Therefore, they cause more inconvenience than the potential benefits they might yield.

\item \textit{Suggestions about conflicting changes must be accurate and scalable}. There exist solutions which attempt to merge changes proactively between a developer's local branch and the latest version of main branch or two developer branches. These tools notify the developers when they detect a merge conflict situation \cite{Palantir, EarlyDetection-1, EarlyDetection-2/2025113.2025139}. Such solutions are impractical to implement in large development environments as the huge infrastructure costs incurred by them may outweigh the gains realized in terms of saved developer productivity. 
\end{enumerate}

Keeping these observations in mind, we propose ConE, a novel technique to i) calculate the Extent Of Overlap (EOO) between two \pr{}s that are active at the same time frame, and ii) determine the existence of Rarely Concurrently Edited files (RCEs). We also derived thresholds to filter out noise and implemented ranking techniques to prioritize conflicting changes.

We have implemented and deployed ConE on \NumRepos\ repositories across different product lines and large scale cloud development environments within Microsoft. Since deployed, in March 2020, ConE evaluated \NumPrs\ \pr{}s and made \NumDecorations\ recommendations about conflicting changes.

This paper describes ConE and makes the following contributions: 
\begin{itemize}
    \item We characterize empirically how concurrent edits and the probability of source code files introducing bugs vary based on the fashion in which edits to them are made, i.e., concurrent vs non-concurrent edits (Section \ref{empiricalstudy}).
    \item We introduce the ConE algorithm that leverages light-weight heuristics such as the extent of overlap and the existence of rarely concurrently edited files, and ConE's thresholding and ranking algorithm that filters and prioritizes conflicting changes for notification (Section \ref{SystemDesign}).
    \item We provide implementation and design details on how we built ConE as a scalable cloud service that can process tens of thousands of \pr{}s across different product lines every week (Section \ref{Sec:Implementation}).
    \item We present results from our quantitative and qualitative evaluation of the ConE system (Section \ref{sec:results}).
\end{itemize}

To the best of our knowledge, this is the first study of an early conflict detection system that is also deployed, in a large scale, cloud based, enterprise setting comprised of a diverse set of developers who work with multiple frameworks and programming languages, on multiple disparate product lines and who are from multiple geographies and cultural contexts. We have observed overwhelmingly positive response to this system with a \PositiveResponseRate\% positive feedback provided by the end users: A very good user interaction rate (2.5 clicks per recommendation that is surfaced by ConE to learn more about conflicting changes) and 93.75\% of the users indicating their intent to use or keep using the tool on a daily basis.

Our interactions and interviews with developers across the company made us realize that developers find it valuable to have a service that can facilitate better communication among them about edits that are happening elsewhere (to the same files or functions that are being edited by them) through simple and non-obtrusive notifications. This is reflected strongly in the qualitative feedback that we have received (explained in detail in section \ref{sec:results}).
\section{Related Work}
The software engineering community has extensively studied the impact of merge conflicts on software quality \cite{articleconflict, 8170085}, investigated various methodologies and tools that can help developers discover conflicting changes through interactive visualizations, and developed speculative analysis tools \cite{BreakingtheCode, IntegratingPrograms, ConfigurationManagementTool}. While ConE draws inspiration from some of this prior work, it is more ambitious, targeting a method that is effective while not resource intensive, can be easily scaled to work on tens of thousands of repositories of all sizes, is easy to integrate and fits naturally into existing software development workflows and tools with very little to no disruption.

A conflict detection system that has to work for large organizations with disparate sets of programming languages, tools, product portfolio and has thousands of developers that are also geographically distributed, has to satisfy the requirements listed below: 
\begin{itemize}
\item \textit{language-independent}: the techniques and tooling built should be language-independent in nature and support repositories that hosts code written in any programming language and should support new languages with no or minimal customization.     
\item \textit{non-intrusive}: the recommendations passed by the tool should naturally fit into developer workflows and environment.
\item  \textit{scalable}: finally, the techniques proposed and the system should be performant and responsive without consuming a lot of computing resources and demanding lot of infrastructure to scale them up. 

We now explain some of the prior work that is relevant and explain why they do not satisfy some or all of the requirements.
\end{itemize}

\textbf{Tools based on edit activity.}
Manhattan \cite{6613849} is a tool that generates visualizations about team activity whenever a developer edits a class and notifies developers through a client program, in real time. While this shows useful 3D visualizations about merge conflicts in the IDE itself (thus being non-intrusive and natural to use), it is not adaptive (it does not automatically reflect any changes to the code in the visualization,  unless  the  user decides to re-import the code base), not generic (it works only for Java and Eclipse) and not scalable as it operates on the client side and has to go through the cycle of import-analyze-present again and again for every change that is made, inside the IDE environment. Similarly, FASTDash \cite{FASTDash} is a tool that scans every single file that is edited/opened in every developer local workspace and communicates about their changes back and forth through a central server. This is impractical to implement across large development teams. It requires tracking changes at the client side with the help of an agent program that runs on each client. Furthermore it then keeps listening to every file edit activity in the workspace, then communicating that information with a central server which mediates communication between different workspaces. This is prone to failures and runs into scale issues even with a linear increase in developers and \pr{}s in the system.

\textbf{Tools based on early merging.} 
Some tools were built upon the idea of attempting actual merging and notifying the developers through a separate program that runs on the client \cite{Palantir, EarlyDetection-1, EarlyDetection-2/2025113.2025139}. These solutions are very resource intensive because the system needs to perform the actual source code merge for every \pr{} or developer branch with the latest version of the main branch (despite implementing optimization techniques like caching and tweaking the algorithm to compute relationships between changes when there is a change to the history of the repository). It is not possible to implement and scale this at a company like Microsoft where tens of thousands of \pr{}s are created every week. Additionally, these solutions do not attempt to merge between two different user branches or two different active \pr{}s but attempt to merge a developer branch with the latest version of the main branch. This will not find conflicting changes that exist in independent developer branches and thus cannot trigger early intervention. Palantir \cite{Palantir} is a tool that addresses some of the performance issues by leveraging a cache for doing dependency analysis.
It is, however, still hard to scale due to the fact that there is client-server communication involved between IDEs and centralized version control servers to scan, detect, merge and update every workspace with information about remote conflicting changes. Some solutions explore speculative merging \cite{10.1145/2025113.2025139, 6606619, 6227180} but the concerns with scalability, non-obtrusiveness remain valid with all of them.

\textbf{Predictive tools.} Owhadi-Kareshk \etal explored the idea of building binary classifiers to predict conflicting changes \cite{8870173}.
Their model consists of nine features, of which the number of jointly edited files is the dominant one.
The model has been evaluated on a dataset of syntactic merge conflicts reverse engineered from git histories.
The model's reported performance in terms of precision ranges from 0.48 to 0.63 (depending on the programming languages).

While one of our proposed metrics, our Extent of Overlap, is akin to the dominant feature in Owhadi-Kareshk's model, unfortunately their proposed approach cannot be applied in our context.
In particular the reported precision is too low and would generate too many false alarms which would render our tool unused \cite{GoogleLessons}.
Furthermore, the reported precision and recall are measured based on a gold standard of syntactic changes.
Instead, we target an evaluation with actual developers, based on a service deployed on repositories they are working with on a daily basis.
As we will see in our evaluation, these developers not only value warnings about syntactic changes, but also 
semantic conflicts \cite{SemanticConflicts}, or even cases of code/effort duplication (as explained in Section \ref{QualitativeAnalysis}).

\textbf{Empirical studies of merge conflicts and collaboration.}
There exists many studies that do not propose tools, but study merge conflicts or present methods to predict conflicts or recommend coordination. Zhang \etal \cite{6385141} conducted an empirical study of the effect of file editing patterns on software quality. They conducted their study on three open source software systems to investigate the individual and the combined impact of the four patterns on software quality. To the best of our knowledge ours is the first empirical study that is conducted at scale, on industry data. We perform analysis on 67K bug reports, from 83K files (in comparison to the studies conducted by Zhange \etal which looked at 98 bugs, from 2,140 files).

Ashraf \etal presented reports from mining cross-task artifact dependencies from developer interactions \cite{8667990}.
Dias \etal proposed methods to understanding predictive factors for merge conflicts \cite{article-Dias}, i.e.,  how conflict occurrence is affected by technical and organizational factors.
Studies conducted by Blincoe \etal and Cataldo \etal \cite{8667990, 10.1145/1180875.1180929} show the importance of timely and efficient recommendations and the implications for the design of collaboration awareness tools. Studies like this form a basis for building solutions that are scalable and responsive (the large-scale ConE service that we deployed at Microsoft) and their importance in creating awareness of the potential conflicts.

Costa \etal proposed methods to recommend experts for integrating changes across branches \cite{10.1145/2950290.2950339} and characterized the problem of developers' assignment for merging branches \cite{article-costa}.
They analyzed merge profiles of eight software projects and checked if the development history is an appropriate source of information for identifying the key participants for collaborative merge. They also presented a survey on developers about what actions they take when they need to merge branches, and especially when a conflict arises during the merge. Their studies report that the majority of the developers (75\%) prefer collaborative merging (as opposed to merging and taking decisions alone). This reiterates the fact that tools that facilitate collaboration, by providing early warnings, are important in handling merge conflict situations.

\section{Concurrent versus non-concurrent edits in practice} \label{empiricalstudy}
The differences in the fashion in which edits are made to source code files (concurrent vs non-concurrent) can cause various unintended consequences (as explained in section \ref{Intro}). We performed large scale empirical analysis of source code edits to understand the ability of concurrent edits to cause bugs. We picked bugs as a candidate for our case study because it is relatively easy to mine and generate massive amounts of ground truth data about bugs and map them back to the changes that induced the bugs, by leveraging some of the techniques proposed by Wang \etal \cite{10.1145/3345629.3345635}, at Microsoft's scale. Understanding the extent of the problem, i.e., the side effects caused by concurrent source code edits in a systematic way, is an essential first step towards making a case for building an early intervention service like ConE. This allows us to quickly sign up customers inside the company and deploy the ConE system on thousands of repositories, for tens of thousands of developers, across Microsoft. To that extent, we formulate two research questions that we would like to find answers for.
\begin{itemize}
    \item \textbf{RQ1} How do concurrent and non-concurrent edits to files compare in the number of bugs introduced in these files?
    \item \textbf{RQ2} To what extent are concurrent, non-concurrent, and all edits, correlated with subsequent bug fixes to these files?
\end{itemize}

Answering the questions above allows us to assess the urgency of the problem. The methods, techniques and outcomes used can also be employed to inform decision makers, when investments in the adoption of techniques like ConE need to be made.

We performed an empirical study on data that is collected from multiple, differently sized repositories. 
For our study, we focused on one of the important side effects that is induced by collaborative software development, i.e., the ``number of bugs introduced by concurrent edits''.
We chose this scenario as we have an option to generate an extensive set of ground truth data, by leveraging techniques proposed by Wang \etal \cite{10.1145/3345629.3345635}, to tag \pr{}s as bug fixes.
They employ two simple heuristics to tag bug fixes: the commit message should contain the words “bug” or “fix”, but not “test case” or “unit test”.
Tagging changes that introduce bugs is not a practice that is followed very well in organizations. Studies have shown that files changed in bug fixes can be considered as a good proxy to files that introduced the bugs in the first place \cite{4019564, 10.1145/1390817.1390826}. Combining both ideas we created a ground truth data set which we used in our empirical analysis. We broadly classify our empirical study into three main steps.
\begin{enumerate}
    \item Data collection: Collect data using the data ingestion framework that we have built which ingests meta data about \pr{}s (author, created/closed dates, commits, reviewers etc), iterations/updates of \pr{}s, file changes in \pr{}s, and intent of the \pr{} (feature work, bug fix, refactoring etc).
    \item Use the data collected in Step 1 to analyze the impact of concurrent edits on bugs or bug fixes in comparison to non-concurrent edits.
    \item Explain the differences in correlations between concurrently versus non-concurrently edited files to the number of bugs that they introduce.
\end{enumerate}
For the purpose of the empirical analysis we define concurrently and non-concurrently edited files as follows:

\begin{itemize}
    \item Concurrently edited files: Files which have been edited in two or more \pr{}s, at the same time, while the \pr{}s are active. A \pr{} is in an `active' state when it is being reviewed but not completed or merged.
    \item Non-concurrently edited files: Files which have never been edited in two \pr{}s while they both are in active state. So, we are sure that changes made to these files are always made in the latest version and are merged before they are edited through another active \pr{}
\end{itemize}

\subsection{Data Collection} \label{DataCollection}
We collected data about file edits (concurrent and non-concurrent) from the \pr{} data, for six months, from six repositories. We picked repositories in which at least 1,000 \pr{}s are created every month. After reducing the repositories to a subset, we randomly selected six repositories for the purpose of the analysis.
We made sure our data set is representative in various dimensions like size (small (1), medium (2), large (3)), the nature of the product (on-prem product (2) vs cloud service (4)), geographical distribution of the teams (US only (2) versus split between different countries and time zones (4)), and programming languages (as listed in Table \ref{FileTypeDistribution}).
We performed data cleansing by applying the filters listed below: 
\begin{itemize}
    \item Exclude PRs that are open for more than 30 days: the majority of these \pr{}s are `Stale PRs' which will be left open forever or abandoned at a later point of time. Studies shows that 70\% of the \pr{}s gets completed within a week after creation \cite{conf/icse/GousiosPD14}.
    \item Exclude PRs with more than 50 files (this is the 90th percentile for file counts in our \pr{} data set). This is one of the proxies that we use to to exclude PRs which are created by non-human developers that do mass refactoring or styling changes etc.
    \item Exclude edits made to certain file types. We are primarily interested in understanding the effects of concurrent edits on source code changes as opposed to files like configuration or initialization files which are edited by lot of developers through lot of concurrent \pr{}s, all the time. For the purpose of this study, we consider only the following file types:  .cs, .c, .cpp, .ts, .py, .java, .js, .sql.
    \item Exclude files that are edited a lot: 
    For example, files that contain global constants, key value pairs, configuration values, or enums are usually seen in a lot of active \pr{}s at the same time.
    We studied 200 \pr{}s to understand the concurrent edits to these files. They typically are in the order of a few thousands of lines in size, which is well above the median file size.
    In all cases the edits are localized to different areas of the files and surgical in nature.
    Sometimes, the line numbers of the edits are far away (few thousands of lines away, at least). Therefore, we impose a filter on the edit count of fewer than twenty times in a month (90th percentile of edit counts for all source code files) and exclude any files that are edited more than this. Without this filter, these frequently edited files would dominate the results of the ConE recommendations thus yielding too many warnings for harmless concurrent edits.
\end{itemize}

\begin{table}
\setlength\belowcaptionskip{7pt}
\caption{Distribution of concurrently and non-concurrently edited files per repository}
 \begin{tabular}{p{1.5cm}|R{1.5cm}|R{2cm}|R{1.5cm}|R{2cm}|R{2cm}}
 \toprule
    \textbf{Repo} & \textbf{Distinct number of concurrently edited files} & \textbf{Distinct number of non-concurrently edited files} &
    \textbf{Number of bug fix \pr{}s} & \textbf{Percentage of concurrently edited files} & \textbf{Percentage of non-concurrently edited files} \\
    \midrule
    Repo-1 & 3500 & 4875 & 4781 & 41.7 & 58.2 \\ 
    Repo-2 & 10470 & 16879 & 15678 & 38.2 & 61.8 \\ 
    Repo-3 & 2907 & 4119 & 5467 & 41.3 & 58.7 \\ 
    Repo-4 & 5560 & 7550 & 8972 & 42.4 & 57.6 \\ 
    Repo-5 & 4110 & 7569 & 9786 & 35.2 & 64.8 \\ 
    Repo-6 & 5987 & 9541 & 9443 & 38.5 & 61.5 \\
    \midrule
    Total & 32534 & 50533 & 54127 & 39.1 & 60.9 \\
    \bottomrule
\end{tabular}
\label{Tab:EditDistribution}
\end{table}

We started with a data set of 208,556 \pr{}s. As bug fixes is our main concentration for the empirical analysis, we removed all the \pr{}s that are not bug fixes. That reduced the data set to 67,155 \pr{}s (32.2\% of the \pr{}s are bug fixes). Then we applied other filters mentioned above, which further reduced the data set to 54,127 \pr{}s (25.95\%). Table \ref{Tab:EditDistribution} shows the distribution of concurrently and non-concurrently edited files per repository.

\subsection{RQ1: Concurrent versus non-concurrent bug inducing edits} \label{DataAnalysis}
We take every (concurrently or non-concurrently) edited file, and check whether the nature of the edit has any effect on the likelihood of that file appearing in bug fixes after the edit has been merged.
We compare how the percentage of edited files that are seen in bug fixes (within a day, a week, two weeks and a month), varies with the nature of the edit (concurrent vs non-concurrent).

\begin{figure*}
\subfloat[\label{fig:a}]
        {\includegraphics[width=0.450\textwidth]{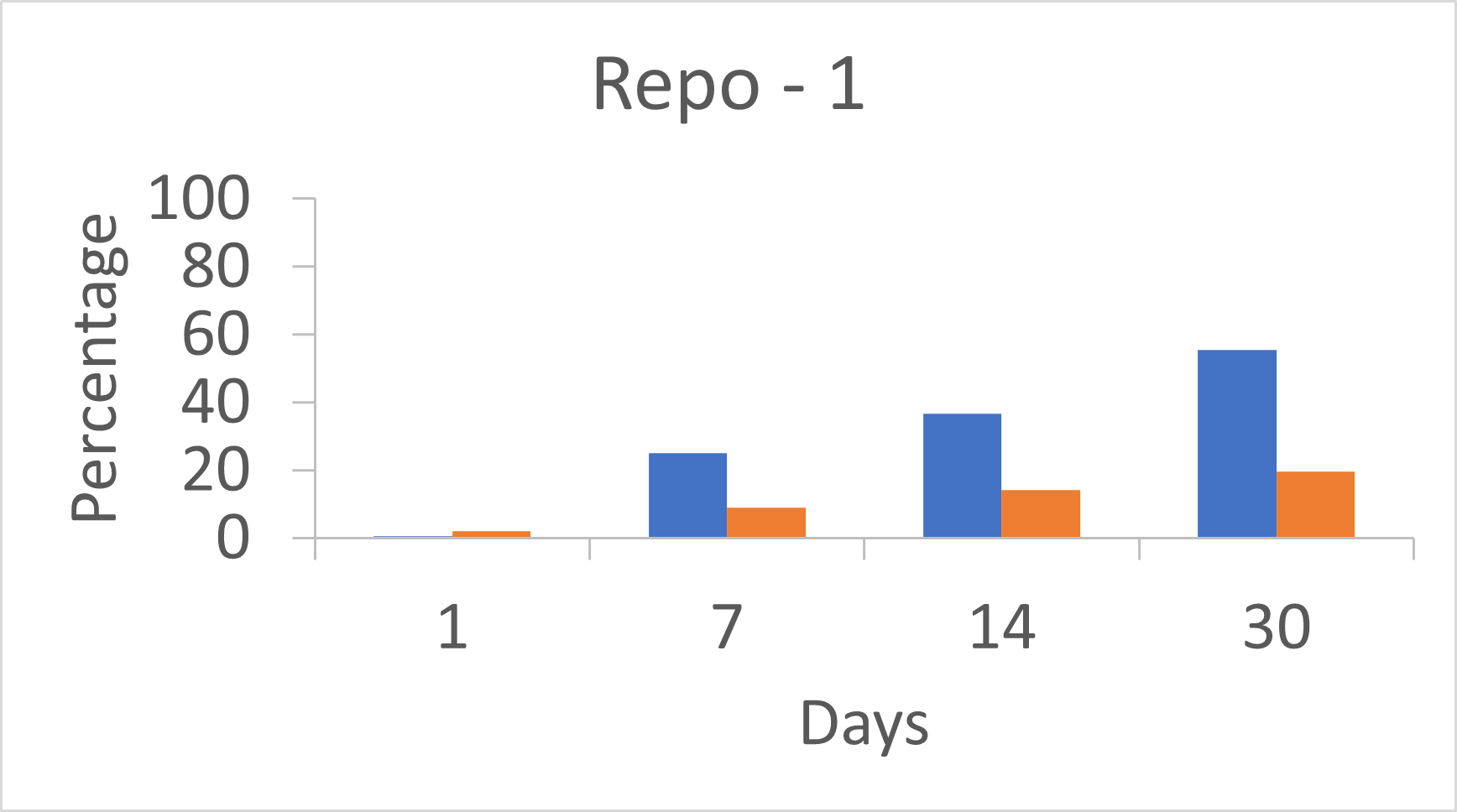}}
    \hfill
\subfloat[\label{fig:b}]
        {\includegraphics[width=0.450\textwidth]{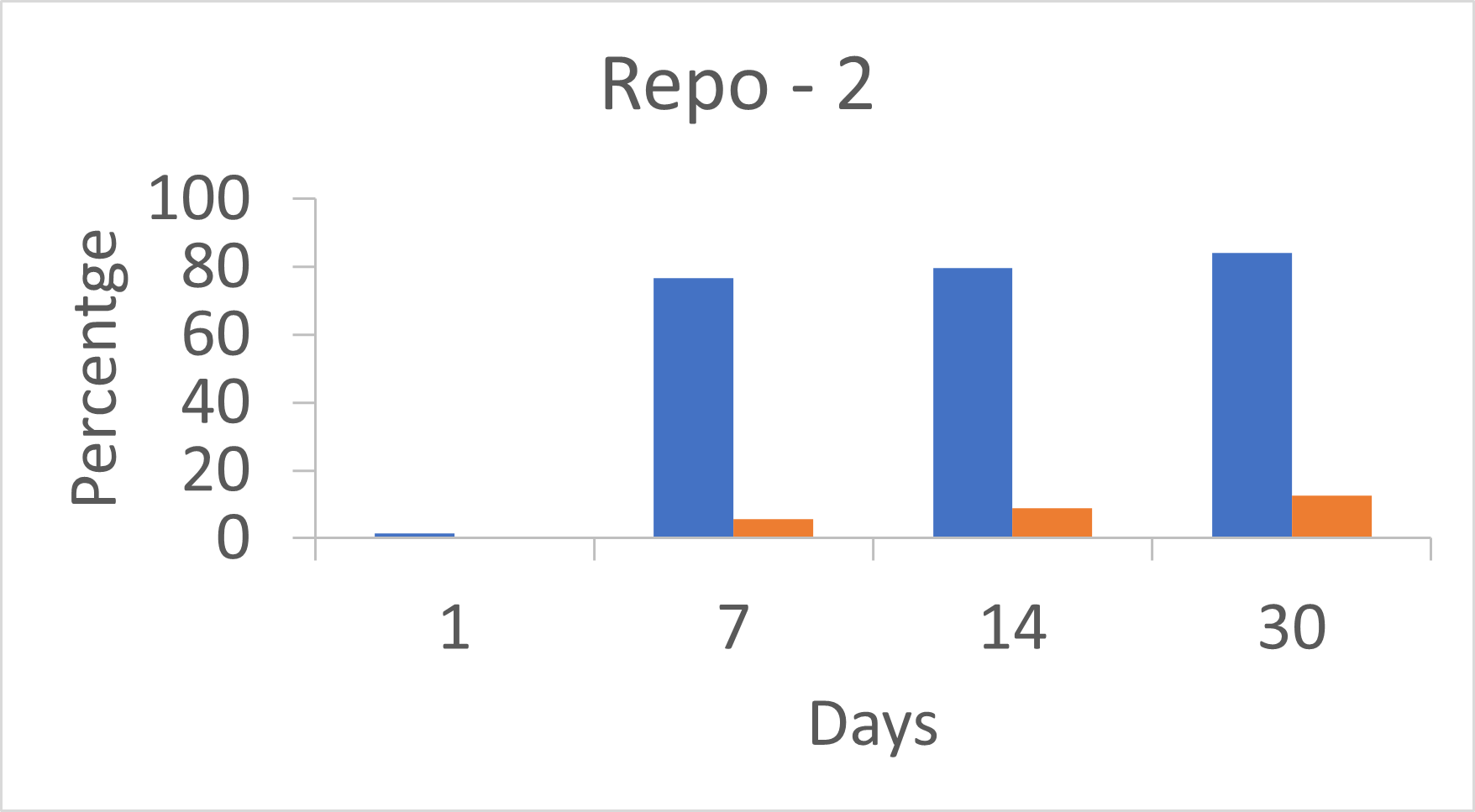}}
    \hfill
\subfloat[\label{fig:c}]
        {\includegraphics[width=0.450\textwidth]{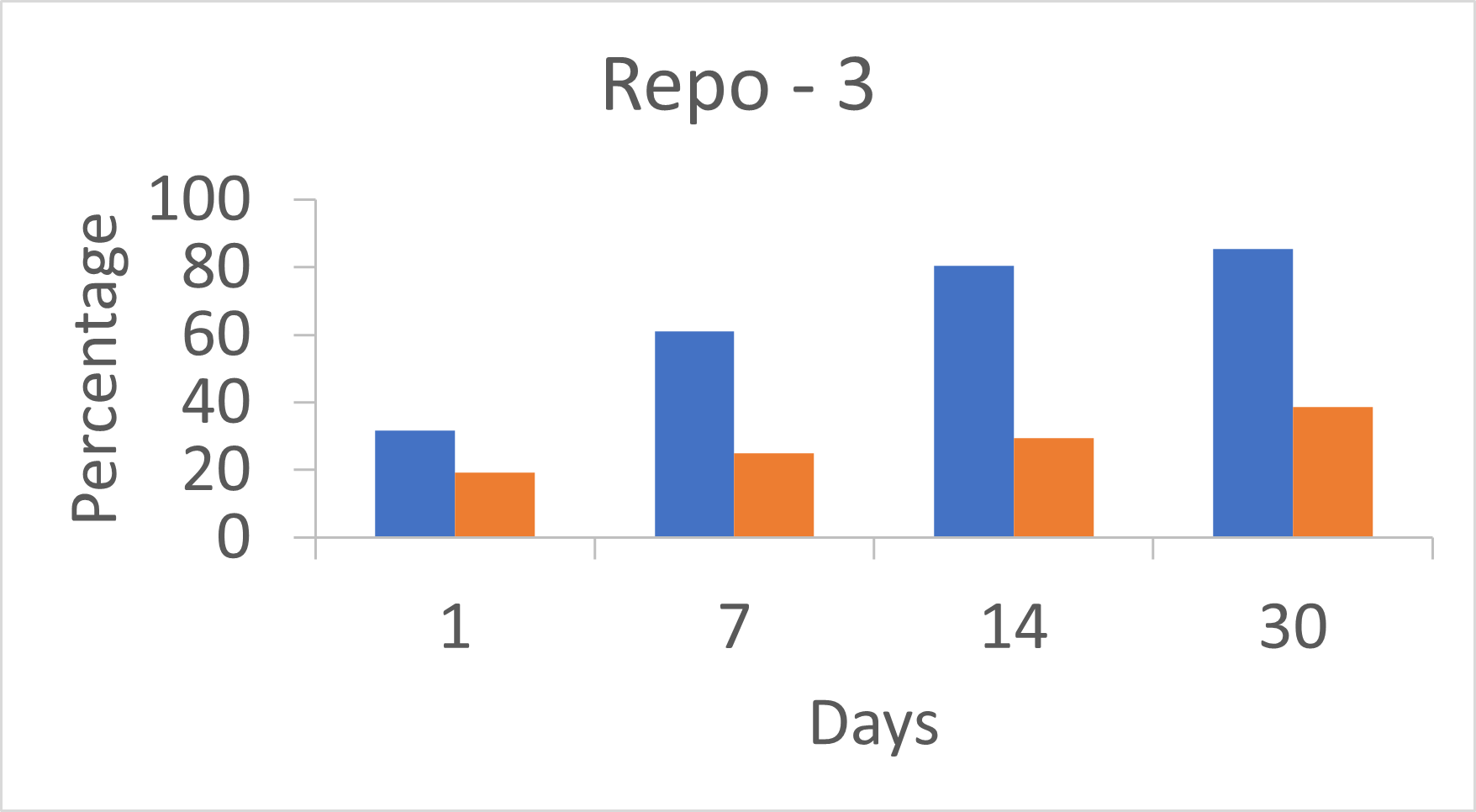}}
    \hfill
\subfloat[\label{fig:d}]
        {\includegraphics[width=0.450\textwidth]{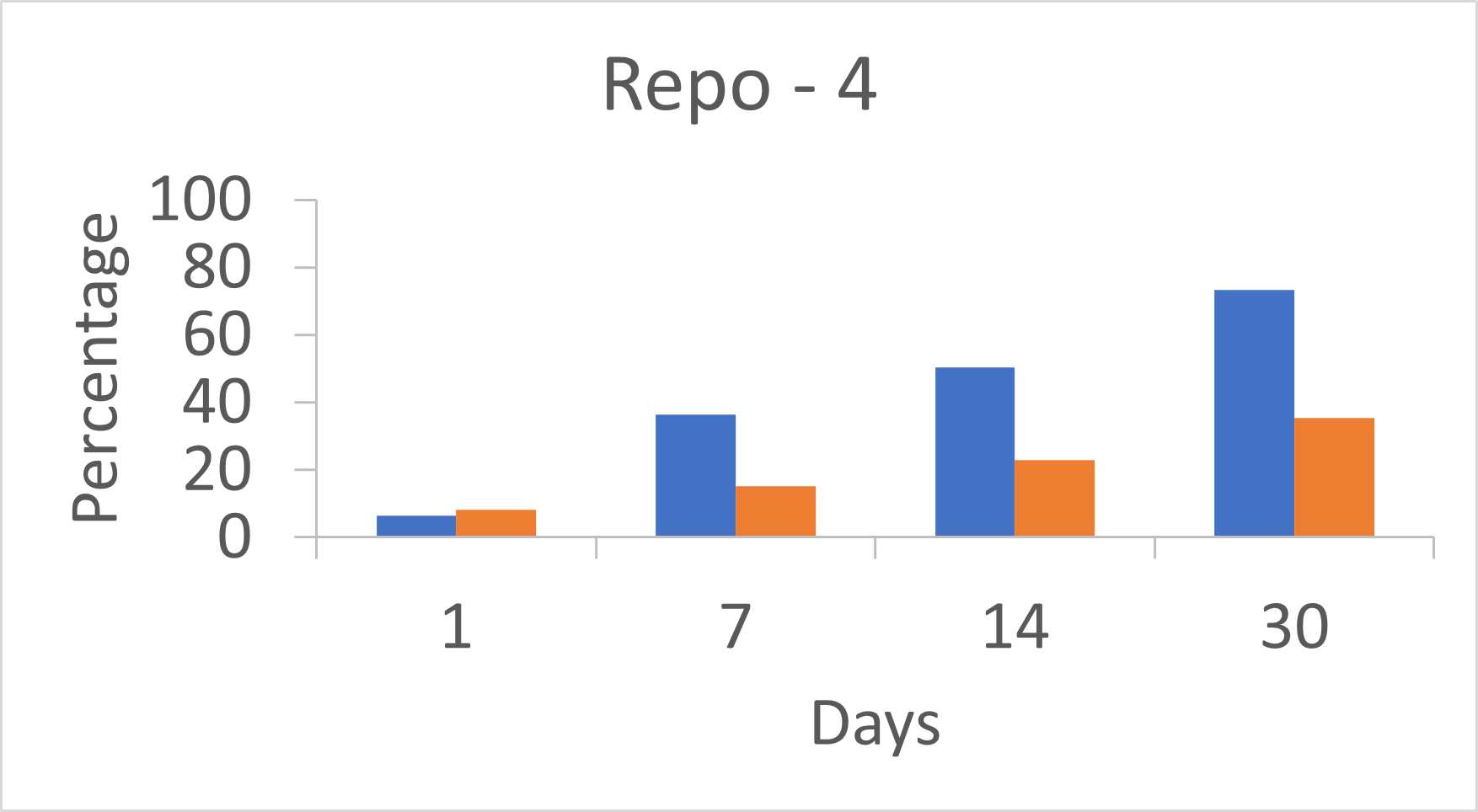}}
    \vskip\baselineskip
\subfloat[\label{fig:e}]
        {\includegraphics[width=0.450\textwidth]{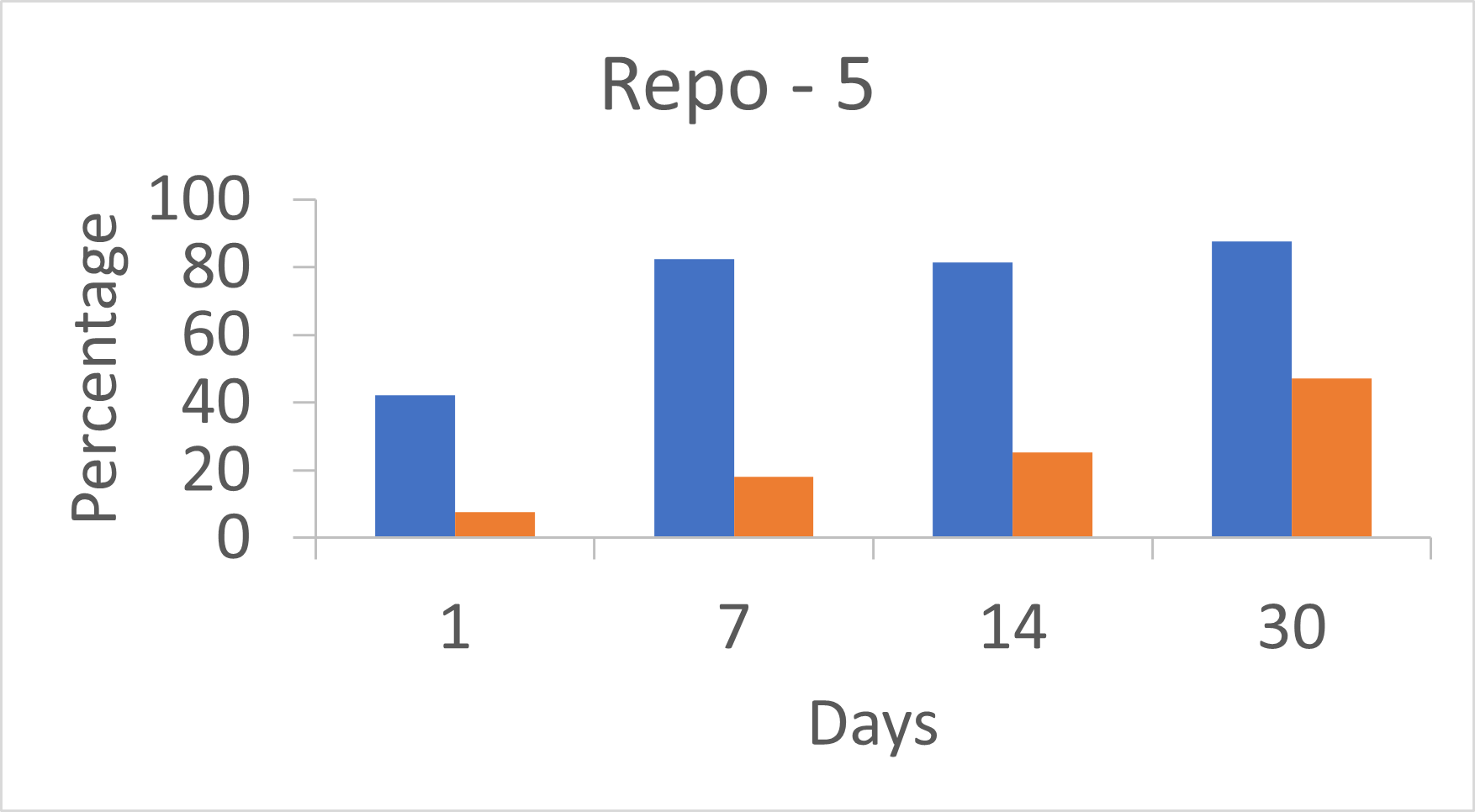}}
    \hfill
\subfloat[\label{fig:h}]
        {\includegraphics[width=0.450\textwidth]{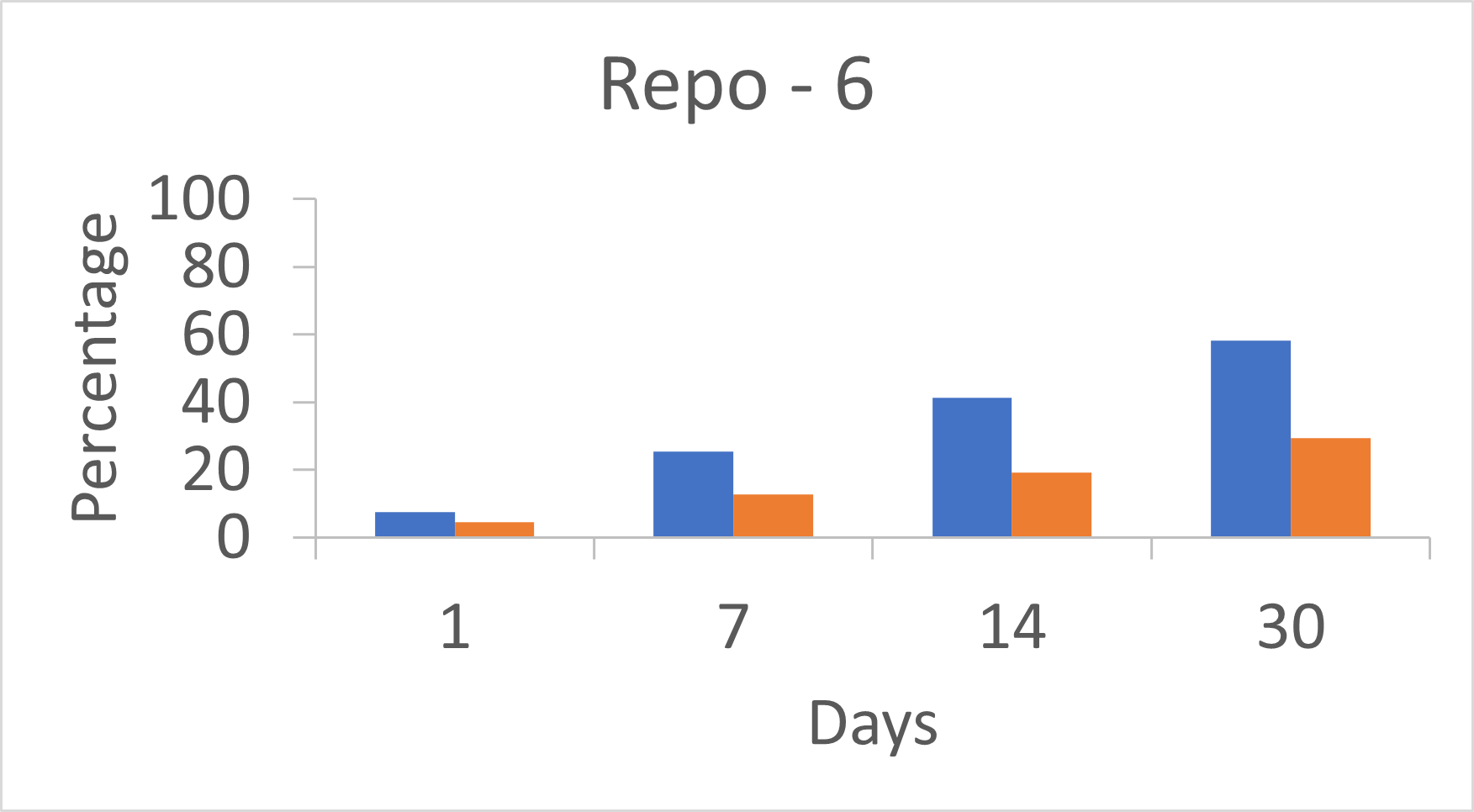}}
    \vskip\baselineskip
\caption{Graphs showing how the percentage of files seen in bug fixes (within a day, a week, two weeks and a month) is changing.
\Description{Concurrent versus non-concurrent bug inducing edits for six repositories}
\textcolor{blue}{Blue} and \textcolor{orange}{orange} bars represent concurrently and non-concurrently edited files, repsectively.}
\label{fig:ConcurrentVsNonConcurrentEdit}
\end{figure*}

Figure \ref{fig:ConcurrentVsNonConcurrentEdit} shows the impact of concurrent versus non-concurrent edits on the number of bugs being introduced.
Across all six repositories, the percentage of bug inducing edits is consistently higher for concurrently edited files (blue bars) than for non-concurrently edited ones (orange bars).

\subsection{RQ2: Edits in files versus bug fixes in files} \label{CorrelationAnalysis}

We use Spearman's rank correlation to analyze how the total number of edits, concurrent edits, and non-concurrent edits to files each correlate with the number of bug fixes seen in those files.

While Figure \ref{fig:ConcurrentVsNonConcurrentEdit} shows that more concurrently edited files are seen in bug fix \pr{}s (compared to non-concurrently edited ones), this might also be because these files are frequently edited and seen in bug fix \pr{}s naturally. To validate this, we performed Spearman rank correlation analysis for each file that is ever edited with respect to how many times it is seen in bug fixes
(the numbers of data points from the six repositories are listed in Table \ref{Tab:EditDistribution}):
\begin{itemize}
    \item The total number of times a file is seen in \emph{all} completed \pr{}s vs the number of bug fixes in which it is seen
    \item The total number of times a file is seen in \emph{concurrent} \pr{}s vs the number of bug fixes in which it is seen
    \item The total number of times a file is seen in \emph{non-concurrent} \pr{}s vs the number of bug fixes in which it is seen
\end{itemize}

The results are in Table \ref{Tab:Correlation-Comparison}.
We observe that concurrent edits (third column) consistently are correlated with bug fixes, more so than non-concurrent edits (column 4) and all edits (column 2).
For all repositories except Repo-4, there exists almost no correlation between non-concurrent edits (column 4) and bug fixes.

For Repo-4, frequently edited files are not necessarily the ones seen in more bug fixes: there exists a \emph{negative} correlation between total edits (column 2) and the number of bug fixes. However, files that are \emph{concurrently} edited (column 3) do have a positive correlation with the number of bug fixes. 

The variety in the correlations can be explained by the fact that concurrent editing is just one of many factors related to the need for bug fixing. Other factors might include the level of modularization, developer skills, the test adequacy, engineering system efficiency, and so on. 

\begin{table}
\centering
\caption{Spearman rank correlation analysis for total edits, concurrent edits, non-concurrent edits vs bug fixes. }
 \begin{tabular}{l|S|S|S}
 \toprule
      Repo &
      {\shortstack{Total Edits to \\ Bug Fixes}} &
      {\shortstack{Concurrent Edits to \\ Bug Fixes}} &
      {\shortstack{Non-Concurrent Edits \\ to Bug Fixes}} \\
\midrule 
    Repo-1 & 0.145{$^{***}$} & 0.298{$^{***}$} & 0.034{$^{**}$} \\ 
    Repo-2 & 0.072{$^{***}$} & 0.140{$^{***}$} & 0.057{$^{**}$} \\ 
    Repo-3 & 0.140{$^{*}$} & 0.330{$^{*}$} & 0.120{$^{*}$} \\ 
    Repo-4 & -0.077{$^{***}$} & 0.451{$^{***}$} & -0.461{$^{***}$} \\ 
    Repo-5 & 0.164{$^{***}$} & 0.472{$^{***}$} & 0.091{$^{***}$} \\ 
    Repo-6 & 0.084{$^{**}$} & 0.196{$^{***}$} & 0.005{$^{*}$} \\ 
    \bottomrule
    \addlinespace[1ex]
\multicolumn{3}{l}{\textsuperscript{***}$p<0.001$, 
  \textsuperscript{**}$p<0.01$, 
  \textsuperscript{*}$p<0.05$}
\end{tabular}
\label{Tab:Correlation-Comparison}
\end{table}
\section{System Design} \label{SystemDesign}
Backed by the correlation analysis suggesting that concurrent edits maybe prone to causing issues. Also, there exists a huge demand from engineering organizations, inside Microsoft, for a better tool that can detect conflicting changes early on and facilitate better communication among developers, we moved forward to materialize the idea of ConE into reality. We then performed large scale testing and validation by deploying ConE on \NumRepos\ repositories.
Details about the implementation, deployment and scale-out are provided in section \ref{Sec:Implementation}.

In this section we describe ConE's conflict change detection methodology, algorithm and system design in detail. We will use the following terminology:

\begin{itemize}
    \item \textit{Reference \pr{}} is a \pr{} in which a new commit/update is pushed thus triggering the ConE algorithm to be run on that \pr{}. 
    
    \item \textit{Active \pr{}} is a \pr{} whose state is `active' when the ConE algorithm is triggered to be run on a reference \pr{}.

\end{itemize}

A key design consideration is that we want to avoid false alarms. In the current state of the practice developers never receive warnings about potentially harmful concurrent edits. Based on this we believe it is acceptable to miss a few warnings. On the other hand, giving false warnings will likely lead to rejection of a tool like ConE. 
For that reason, ConE has several built-in heuristics that are aimed at reducing such false alarms.

Due to the nature of the problem, the domain we are operating in, and the algorithm we have in-place, it is possible to see notifications that are false alarms. One of the design choices that we had to make was to minimize the false alarms by making it more conservative. A side effect of this is our coverage (number of \pr{}s for which we send a notification) will be lower. Studies have shown that, in large organizations, tools that generate many false alarms are not used and eventually deprecated \cite{GoogleLessons}. However, recent techniques proposed by Brindescu \etal \cite{inproceedings-brindescu}, can potentially aid in facilitating a decision by determining the merge conflict situations to flag, based on the complexity of the merge conflict.  

\subsection{Core Concepts} \label{DesignOverview}
ConE constantly listens to events that happen in an Azure DevOps environment \cite{AzDo}. When any new activity is recorded (e.g., pushing a new update or commit) in a \pr{}, the ConE algorithm is run against that \pr{}. Based on the outcome, ConE notifies the author of the \pr{} about conflicting changes. We describe two novel constructs that we came up with for detecting conflicting changes and determining candidates for notifications: Extent of overlap (EOO) and the existence of `Rarely Concurrently Edited' files (RCEs). Next, we provide a detailed description of ConE's conflict change detection algorithm and the parameters we have in place to tune ConE's algorithm.

\subsubsection{Extent of Overlap (EOO)} ConE scans all the active \pr{}s which meet our filtering criteria (explained in section \ref{DataCollection}) and for each such \pr{} (reference \pr{}) calculates the percentage of files edited in the reference \pr{} that overlap with each of the active \pr{}s. 
\[ \textit{Extent of Overlap} = \frac{\mid F_r \cap F_a - F_e\mid} {\mid F_r \mid} * 100\]
where F\textsubscript{r} = Files edited in reference \pr{}, F\textsubscript{a} = Files edited in a given active \pr{}, F\textsubscript{e} = Files excluded i.e., files that are not of types listed in the paragraph below.
The idea is to find the percentage of items that are commonly edited in multiple active \pr{}s and create a pairwise overlap score for each of the active and reference \pr{} pairs. Intuitively, if the overlap between two active \pr{}s is high, the probability of them doing duplicate work or causing merge conflicts when they are merged is also going to be high. We use this technique to calculate the overlap in terms of number of overlapping files for now. This can be easily extended to calculate the overlap between two active \pr{}s in terms of number of classes or methods or stubs if that data is available. 

A milder version of EOO is used by the model proposed by Owhadi-Kareshk et al \cite{8870173}, which looks at the number of files that are commonly edited in two \pr{}s when determining conflicting changes. While calculating extent of overlap it is important to exclude edits to certain file types whose probability of inducing conflicts is minimal. This helps in reducing false alarms in our notifications significantly. Based on a manual inspection of 500 randomly selected bug fix \pr{}s, by the first three authors, we concluded that concurrent edits to initialization or configuration files are relatively safe, but that concurrent edits to source code files are more likely to lead to problems. Therefore, we created an \emph{allow list} based on file types as shown in Table \ref{FileTypeDistribution}. As can be seen, this eliminates around 6.4\% of the files. Note that such an \emph{allow list} is programming language-specific. When ConE is to be applied in different contexts, different allow lists are likely needed.

\begin{table}
\caption{Distribution of file types seen in bug fixes}
 \begin{tabular}{l|r|l} 
\toprule 
File type & Percentage & On ConE \emph{allow list}?\\
\midrule
.cs & 44.32 & yes \\
.cpp & 18.55 & yes \\
.c & 11.27 & yes \\
.sql & 6.20 & yes \\
.java & 5.36 & yes \\
.js & 3.98 & yes \\
.ts & 3.79 & yes \\
.ini & 0.20 & no \\
.csproj & 0.04 & no \\
others & 6.29 & no \\
\bottomrule
\end{tabular}
\label{FileTypeDistribution}
\end{table}

\begin{table}
\caption{Number of Bug Fixes with RCEs and No RCEs}
 \begin{tabular}{l|r|r} 
\toprule 
Edit type & Count & Percentage \\
\midrule
Bug fix PRs with no RCEs & 1617 & 78.3 \\ 
Bug fix PRs with at least one RCE & 446 & 21.7 \\
\bottomrule
\end{tabular}
\label{RCE-comparison}
\end{table}

\subsubsection{Rarely Concurrently Edited files (RCEs)} These are the files which typically are \textit{not edited concurrently}, recently. Usually all the updates or edits to them are performed, in a controlled fashion, by a single person or small set of people. Seeing RCEs in multiple active \pr{}s is an anomalous phenomenon. For example, a file foo.cs is always edited by a given developer, through one active \pr{} at any point. The ConE system keeps a track of such files and tags them as RCEs. In the future, if multiple active \pr{}s are seen editing this file simultaneously, ConE flags them. Our intuition is that, if a lot of RCEs are seen in multiple active \pr{}s, which is unusual, changes to these files should be reviewed carefully and everyone involved in editing them should be aware of others' changes. 

We performed an empirical analysis, from our shadow mode deployment data (as explained in Section \ref{Deployment}), to understand how pervasive RCEs really are. As explained in Table \ref{RCE-comparison}, 21.7\% of bug fixes contains at least one RCE in them while the total number of RCEs in these repositories is \textit{just ~ 2\%}. Based on this data and anecdotal feedback from developers, we realized that concurrent edits to RCEs is an unusual activity which should not be seen a lot. But, if observed, it should be notified to all the developers involved.

For building the ConE system, we ran the RCE detection algorithm that looks at the \pr{}s that are created in a repository within the last three months from when the algorithm runs. The duration can be increased or decreased based on how big or how active the system is. This process, after each run, creates a list of RCEs. Once the initial bootstrapping is done and a list of RCEs is prepared, that list is used by the ConE algorithm when checking for the existence of the RCEs in a pair of \pr{}s. The RCE list is updated and refreshed once every week, through a separate process. The process of detecting and updating RCEs is resource intensive. So, we need to strike a balance between how quickly we would like to update the RCE list versus how many resources we need to throw at the system, without compromising the quality of the suggestions. We picked one week as the refresh interval through multiple iterations of experiments. This process guarantees that the ConE system reacts to the changes in the rarity of concurrent edits, especially the cases where an RCE becomes a non-RCE due to the concurrent edits it experiences. The steps involved in creating and updating RCEs are listed below.

Creating the RCE list:
\begin{enumerate}
    \item Get all the \pr{}s created in the last three months from when the algorithm is run. Create a list of all the files that are edited in these \pr{}s by applying the filters explained in the paragraph above on file types.
    \item Prepare sets of \pr{}s that overlap with others. Prepare a list of files edited in the overlapping \pr{}s by applying the filters explained in the paragraph above on file types.
    \item The list of files created in step-1 minus the list of files created in step-2 constitutes the list of rarely concurrently edited files (RCEs).
\end{enumerate}
    
Updating the RCE list:
\begin{enumerate}[resume]
    \item Remove files from the RCE list if they are seen in overlapping \pr{}s when the algorithm is run the next time. Because, if they are seen in overlapping \pr{}s, they will not be qualified to be RCEs anymore.
    \item Refresh the list by adding the new RCEs discovered in the latest edits, when the algorithm is run again.
\end{enumerate}

\subsection{The ConE Algorithm} \label{ConEAlgo}
ConE’s algorithm to select candidate \pr{}s that developers need to be notified about primarily leverages the techniques explained above: Extent of Overlap (EOO) and existence of Rarely Concurrently Edited files (RCEs).
Together these serve to reduce the total number of active \pr{}s under consideration, in order to pick the \pr{}s that need to be notified about.
The ConE algorithm consists of seven steps listed below:

\textbf{Step 1}: Check if the reference \pr{}'s age is more than 30 days. Studies have shown that \pr{}s which are active for so long may not even be completed \cite{conf/icse/GousiosPD14}. Exclude all such \pr{}s. 

\textbf{Step 2}: Construct a list of files that are being edited in the reference \pr{}. While constructing this set, we exclude any files of types that are not in the allow list from Table \ref{FileTypeDistribution}.

\textbf{Step 3}: Construct a set of files that are being edited in each of the active \pr{}s, using the methodology mentioned in Steps~1 and~2. One extra filter that we apply here is to exclude PRs which are being interacted by the author of the reference \pr{}. 
If the author of the reference \pr{} is already aware of this \pr{} there is no need to notify thems.

\textbf{Step 4}: Calculate the extent of overlap using the formula described in section \ref{DesignOverview}.
For every pair of reference \pr{} PR\textsubscript{r} and active \pr{} PR\textsubscript{a1},
calculate the tuple T\textsubscript{ea1} = \textlangle{}PR\textsubscript{r}, PR\textsubscript{a1}, E\textsubscript{1}\textrangle{},
where E\textsubscript{1} is the extent of overlap between the two pull requests.
Do this for all the active \pr{}s with respect to a reference \pr{}.
At the end of this step we have a list of tuples, T\textsubscript{ea} = [\textlangle{}PR\textsubscript{1}, PR\textsubscript{7}, 55\textrangle{}, \textlangle{}PR\textsubscript{1}, PR\textsubscript{12}, 95\textrangle{}, \textlangle{}PR\textsubscript{1}, PR\textsubscript{34}, 35\textrangle{}....].

\textbf{Step 5}: Check for the existence of rarely concurrently edited files (RCEs) and the number of RCEs between each pair of reference and active \pr{}.
Create a tuple T\textsubscript{r} = \textlangle{}PR\textsubscript{r}, PR\textsubscript{a1}, R\textsubscript{1}\textrangle{} where PR\textsubscript{r} is the reference \pr{}, PR\textsubscript{a1} is active \pr{} and R\textsubscript{1} is the number of RCEs in the overlap of reference and active \pr{}s.
Do this for all reference and active \pr{} combinations.
At the end of this step we have a list of tuples, T\textsubscript{ra} = [\textlangle{}PR\textsubscript{1}, PR\textsubscript{7}, 2\textrangle{}, \textlangle{}PR\textsubscript{1}, PR\textsubscript{12}, 2\textrangle{}, \textlangle{}PR\textsubscript{1}, PR\textsubscript{34}, 9\textrangle{}....]

\textbf{Step 6}: Apply thresholds on the values for extent of overlap and the number of RCEs, as explained in section \ref{knobs}.
For example, we can apply a threshold that we select the \pr{}s whose extent of overlap is greater than 50\% OR there should be at least two RCEs. We go through the list of tuples that we have generated in Steps~4 and~5 above and apply the thresholding criteria. 

\textbf{Step 7}: Apply a ranking algorithm to prioritize the \pr{}s that need to be looked at first if multiple \pr{}s are selected by the algorithm. We rank candidate \pr{}s based on the number of RCEs present and then by the extent of overlap. This is because RCEs being edited through multiple active \pr{}s is an anomalous phenomenon which needs to be prioritized.

\subsection{Default Thresholds and Parameter Tuning} \label{knobs}

In this section we describe the thresholding criteria, and the rationale that needs to be applied while choosing parameter values for large scale deployment. The parameters that we have in place are: the extent of overlap (EOO), the number of rarely concurrently edited files (RCEs), the window of time period  (i.e., the number of months to consider for determining RCEs), and the total number of file edits in the reference PR.

In line with our objectives, we are searching for parameter settings that find actual conflicts, yet minimize false alarms. Furthermore, we target settings that are easy to explain (e.g., ``this PR was flagged because half of the files changed it are also touched in another PR'').

\begin{table}
\caption{Distribution of extent of overlap (EOO)}
 \begin{tabular}{p{3cm}|R{2cm}} 
 \toprule
Percentage of overlap (range) & Number of PRs \\
\midrule
\ \ 0-10 & 309 \\ 
11-20 & 223 \\ 
21-30 & 137 \\ 
31-40 & 87 \\ 
41-50 & 25 \\
51-60 & 359 \\ 
61-70 & 92 \\ 
71-80 & 21 \\ 
81-90 & 23 \\ 
91-100 & 378 \\
\bottomrule
\end{tabular}
\label{fig:EOADistribution}
\end{table}

\paragraph{Threshold for EOO:}
For Extent of Overlap, we explored what would happen if we put the threshold at 50\%:
if at least half of the files edited in another \pr{}, consider it for notification.
To assess the consequences of this, we randomly selected 1654 \pr{}s, which have at least one file overlapped with another \pr{}.
This data set is a subset of the data collected to perform empirical analysis on concurrent edits (see Section \ref{empiricalstudy}). We manually inspected each of these 1654 \pr{}s to make sure the overlap we observe is indeed correct. Our empirical analysis (see Table \ref{fig:EOADistribution}), shows that 50\% of the \pr{}s have an overlap of 50\% or less.
Thus, this simple heuristic eliminates half of the candidate \pr{}s for notification,
substantially reducing potential false alarms, and keeping the candidates that are more likely to be in conflict.

\begin{figure}
\centering
\begin{minipage}{.45\textwidth}
  \centering
\includegraphics[width=1\linewidth]{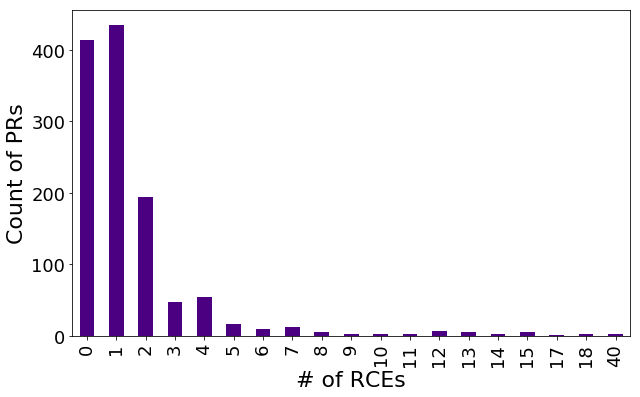}
\caption{Distribution of the number of PRs with a given number of RCEs}
\Description{400 pull requests have no RCEs; another 400 have just one; About 200 have two, and very few PRs have more than 3 RCEs.}
\label{fig:RCEDistribution}
\end{minipage}
\hspace{.5cm}
\begin{minipage}{.45\textwidth}
  \centering
\includegraphics[width=1\linewidth]{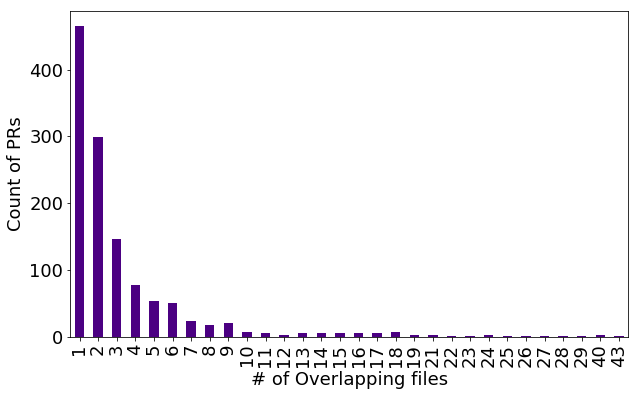}
\caption{Distribution of the number of PRs with a given number of overlapping files.}
\Description{450 pull requests have files that occur in just one PR; 300 pull requests contain files that occur in two PRs; 150 in three or more; then dropping sharply.}
\label{fig:NumFilesistribution}
\end{minipage}
\end{figure}

\paragraph{Threshold for RCEs:}
For RCEs we again followed a simple rule: if the active-reference \pr{} pair contains at least two files that are modified in them, which are always edited in isolation, select the active \pr{} as a candidate.
As shown in Figure \ref{fig:RCEDistribution}, the majority of the \pr{}s contains fewer than two RCEs.
To be conservative, we imposed a threshold on RCE~$\geq$~2, i.e., to select a PR as a candidate, that \pr{} needs to have at least two RCEs that are commonly edited between the reference and active \pr{}s.

\paragraph{Number of overlapping files:}
Assume a developer creates a \pr{} by editing two files and one of them is also edited in another active \pr{}. Here EOO is 50\%.
This means this \pr{} qualifies to be picked as a candidate for notification.
Editing just one file in two active \pr{}s might not be enough to reasonably make an assumption about the potential of conflicts arising.
Therefore, we impose a threshold on the ``number of files'' that needs to be edited, simultaneously, in both \pr{}s.
As a starting point, we imposed a threshold of two, i.e., every candidate \pr{} should have more than two overlapping files (in addition to satisfying the EOO condition of >= 50\%).
We plotted the distribution of the number of overlapping files in Figure \ref{fig:NumFilesistribution}. As shown in Figure \ref{fig:NumFilesistribution}, the number of PRs (on the Y-axis) drops sharply after the number of overlapping file edits is two. Therefore, we picked two as the default threshold.

\paragraph{Threshold Customization:}
In addition to the empirical analysis, we collected initial feedback from developers working with the production systems through our shadow mode deployment (Section~\ref{Deployment}).
One of the prominent requests from the developers was to enable the repository administrators to change the values of the parameters explained above based on the developer feedback.
Therefore, we provided customization provisions to make ConE system suit each repository's needs.
Based on the \pr{} patterns and needs of the repository, system administrators can tune the thresholds to optimize the efficacy of the ConE system for particular reporsitories.

\section{Implementation and deployment} \label{Sec:Implementation}

\subsection{Core Components and Implementation}

\begin{figure}
\begin{center}
\includegraphics[width=.75\hsize]{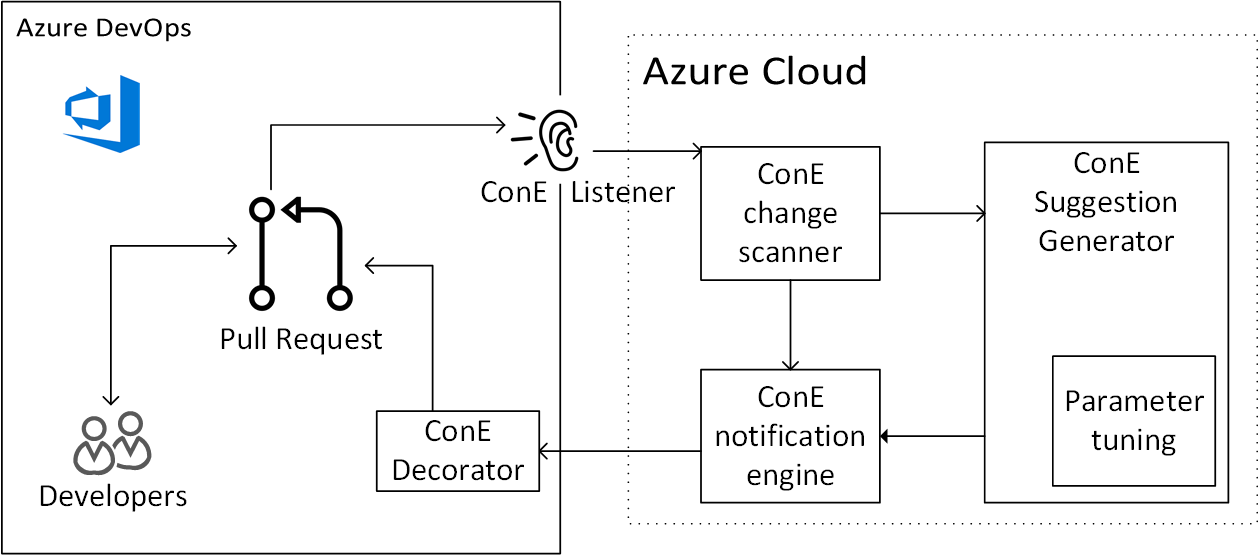}
\end{center}
\caption{\cone System design: Pull requests from Azure DevOps are listened to by the \cone change scanner, suggestion generator, notification engine, and decorator}
\Description{Pull requests from Azure DevOps are listened to by the \cone change scanner, suggestion generator, notification engine, and decorator}
\label{fig:SystemDesign}
\end{figure}

The core \cone components are displayed in Figure~\ref{fig:SystemDesign}.
ConE is implemented on Azure DevOps (ADO), the DevOps platform provided by Microsoft.
We chose to develop ConE on ADO due to its extensibility that allows third party services to interact with \pr{}s through various collaboration points such as adding comments in \pr{}s, a rich set of APIs provided by ADO to read meta data about \pr{}s, and service hooks which allow a third party application to listen to events such as updates that happen inside the \pr{} environment.

Within Azure DevOps, as shown in the left box of Figure~\ref{fig:SystemDesign}, \cone listens to events triggered by pull requests, and has the ability to decorate pull requests with notifications about potentially conflicting other pull requests.
The \cone service itself, shown at the right in Figure~\ref{fig:SystemDesign}, runs within the Azure Cloud.
The \cone change scanner listens to pull request events, and dispatches them to workers in the \cone suggestion generator.
Furthermore, the scanner monitors telemetry data from interactions with \cone notifications.
The core \cone algorithm is offered as a scalable service in the Suggestion Generator,
with parameters tunable as explained in Section~\ref{knobs}.

The ConE Service is implemented using C\# and .NET 4.7. It has been built on top of Microsoft Azure cloud services: Azure Batch ~\cite{AzureBatch} for compute, Azure DevOps service hooks for event notification, Azure worker roles and its service bus for processing events, Azure SQL for data storage, Azure Active Directory for authentication and Application Insights for telemetry and alerting.

\subsection{ConE Deployment} \label{Deployment}

We selected \NumRepos\ repositories to pilot ConE in the first phase.
Some of the key attributes based on which the repository selection process has taken place are listed below:
\begin{itemize}
    \item Prioritize repositories where we have developers and managers who volunteered to try ConE, since we expect them to be willing to provide meaningful feedback.
    \item Include repositories that are of different sizes (based on the number of files present in them): very large, large, medium, and small.
    \item Include repositories that host source code for diverse set of products and services. That includes client side products, mobile apps, enterprise services, cloud services, and gaming services.
    \item Consider repositories which have cross-geography and cross-timezone collaborators, as well as repositories that have most of the collaborators from a single country.
    \item Consider repositories that host source code written in multiple programming languages including combinations of Java, C\#, C++, Objective C, Swift, Javascript, React, SQL etc.
    \item Include repositories which contain a mix of developers with different levels of experience (based on their job titles): Senior, mid-level and junior. 
\end{itemize}

We enabled ConE in \emph{shadow mode} on 60 repositories for two months (with a more liberal set of parameters to maximize the number of suggestions we generate). In this mode we actively listen to \pr{}s, run the ConE algorithm, generate suggestions, and save all the suggestions in our SQL data store for further analysis, without sending the notifications to the developers. We generated and saved 1200 suggestions by enabling ConE in this mode for two months. We then went through the suggestions and the telemetry collected to optimize the system before a large scale roll out. 

The primary purpose of shadow mode deployment is to validate whether operationalizing a service like \cone is even possible at the scale of Microsoft.
Furthermore, it allowed us to check whether we indeed can flag meaningful conflicting pull requests,
and what developers would think of the corresponding notifications.
The telemetry we collected includes the time it takes to run the ConE algorithm, resource utilization, the number of suggestions the ConE system would have made, etc.
We experimented with tuning our parameters (explained in subsection \ref{knobs}) and their impact on the processing time and system utilization. This helped us in understanding the scale and infrastructure requirements and overall feasibility.

\begin{figure*}
\begin{center}
\includegraphics[width=.7\columnwidth]{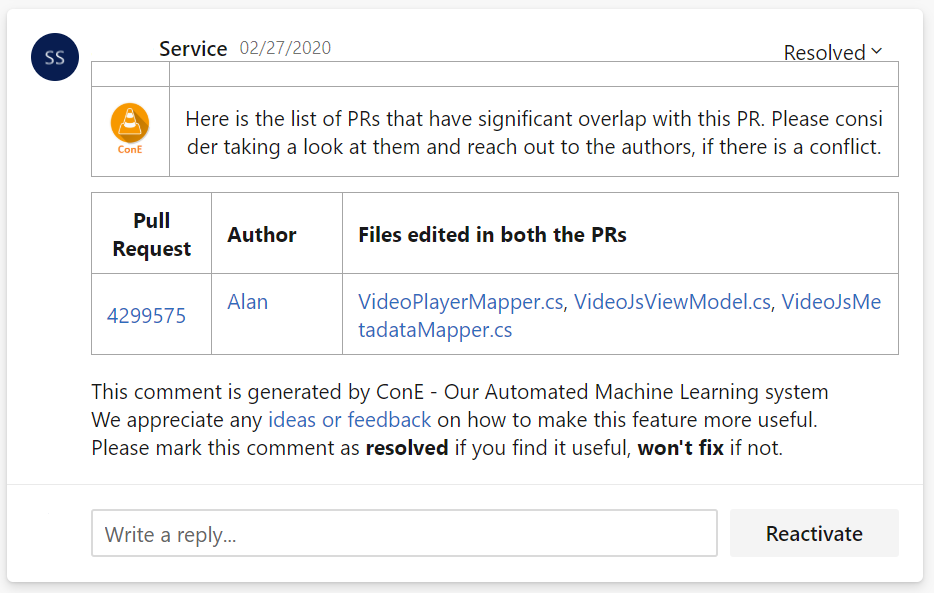}
\end{center}
\caption{\cone notificationt that a \pr has significant overlap with another \pr.}
\Description{Comment generated by \cone listing another pull request that has significant overlap, the author of that pull request, and the files edited in both pull requests.}
\label{fig:Screenshot}
\end{figure*}

We collected feedback from the developers by reaching out to them directly. We have shown them the suggestions we would have made if the ConE system was enabled on their \pr{}s, format of the suggestions and the mode of notifications. We iterated over the design of the notification based on the user feedback before settling on the version of the notification as shown in Figure \ref{fig:Screenshot}.

After multiple iterations of user studies and feedback collection, on the design, frequency, and the quality of the ConE suggestions as validated by the developers participated in our shadow mode deployment program, we turned on the notifications on \NumRepos\ repositories. 

\subsection{Notification Mechanism} \label{Notifications}
We leveraged Azure DevOps's collaboration points to send notifications to developers.
A notification is a comment placed by our system in Azure DevOps \pr{}s.
Figure \ref{fig:Screenshot} shows a screenshot of a comment placed by ConE on an actual \pr{}.
It displays the key elements of a ConE notification: 
a comment text which provides a brief description of the notification,
the id of the conflicting \pr{}, the name(s) of the author(s) of the conflicting \pr{}, files that are commonly edited in the \pr{}s, a provision to provide feedback by resolving or not fixing a comment (marked as ``Resolved'' in the example), and the option to reply to the comment inline to provide explicit written feedback.

While ConE actively monitors every commit that is being pushed to a \pr{}, it will only add a second comment on the same \pr{} again if the state of the active or the reference \pr{} is significantly changed in subsequent updates and ConE finds a different set of \pr{}s as candidates for notification.

In a ConE comment, elements like \pr{} id, file names, author name are actually hyperlinks. The \pr{} id hyperlink points to the respective \pr{}'s page in Azure DevOps. The file name hyperlink points to a page that shows the diff between the versions of the file in the current and conflicting \pr{}s. The author name element, upon clicking, spins up a chat window with the author of the conflicting \pr{} instantly. When people interact with these elements by clicking them, we track those telemetry events (which is consented by the users of the Azure DevOps system, in Microsoft) to better understand the level of interaction developers are having with the ConE system.

\subsection{Scale}
\label{sec:scale}
The ConE system has been deployed on \NumRepos\ repositories in Microsoft.
The repositories have been picked based to maximize the diversity and variety of the repositories in various dimensions as explained in Section~\ref{Deployment}.
Since enabled in March 2020, until September 2020 (when we pulled the telemetry data) ConE evaluated \NumPrs\ \pr{}s which were created in all the repositories on which ConE has been enabled. Within these \NumPrs\ \pr{}s, an additional 156,000 update events (commits on the same branch, possibly affecting new files) occurred.
Thus, ConE had to react to and process a total of 182,000 events that were generated, within Azure DevOps, in those six months. For every update, ConE has to compare the reference \pr{} with all active \pr{}s that match ConE's filtering criteria. In total ConE made a total of approximately two million comparisons.

The scale of operations and processing is expected to grow as we onboard new and large repositories.
The simple and lightweight nature of the ConE algorithm combined with the scalable architecture and efficient design, and its engineering on Azure cloud has given us the ability to process events at this scale with a response rate of less than four seconds per event.
The time it takes to process an event end to end, i.e., receiving the \pr{} creation or update event, running the ConE algorithm and passing the recommendations back (if any) has never taken more than four seconds. ConE employed a single service bus queue and four worker roles in Azure to handle the current scale. As per our monitoring and telemetry (resource utilization on Azure infrastructure, processing latency, etc.) ConE still had bandwidth left to serve the next hundred repositories of similar scale with the current infrastructure setup.

\section{Evaluation: Developers perceptions about ConE's usefulness} \label{sec:results}

Out of the \NumPrs\ \pr{}s under analysis during \cone's six month deployment (Section~\ref{sec:scale}), ConE's filtering algorithm (Section~\ref{ConEAlgo}) excluded 2,735 \pr{}s. In the remaining 23,265 \pr{}s, ConE identified \NumDecorations\ \pr{}s to send notifications to (3.33\%).
In this section, we evaluate the usefulness of these \NumDecorations notifications.

\begin{figure}
\includegraphics[width=1\columnwidth]{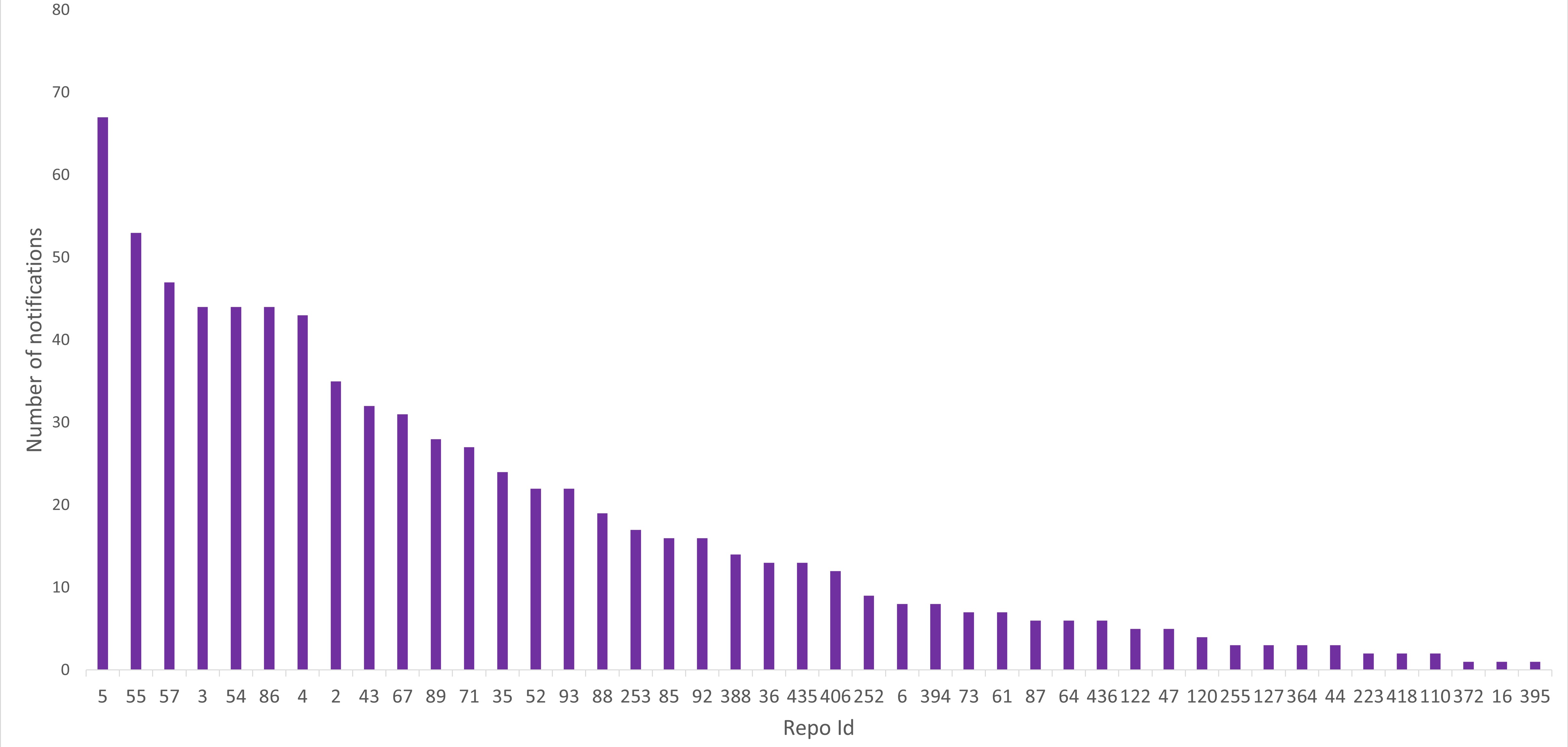}
\caption{Distribution of notifications per repository}
\Description{Two repositories have over 50 notifications, five have betwewwen 40 and 50 notifications, another 20 have between 20 and 40 notifications, and most repositories have 20 or fewer notifications}
\label{Repo-Recommendation}
\end{figure}

All repositories were analyzed with the standard configuration; No adjustments were made to the parameters. Though the service is enabled to send notifications in \NumRepos\ repositories, during the six-month observation period, \cone raised alerts on just 44 distinct repositories. As shown in Figure \ref{Repo-Recommendation}, the notification volume varies between repositories. 

\subsection{Comment resolution percentage}
\cone offers an option for users to provide explicit feedback on every comment it placed, within their \pr{}s.
Users can select the ``Resolved'' option if they like or agree with the notification, and the ``Won't fix'' option if they think it is not useful.
A subset of users were given instructions and training on how to use these options.
The notification itself also contains instructions, as shown in Figure~\ref{fig:Screenshot}.
A user can choose not to provide any feedback by just leaving the comment as is, in the ``Active'' state.
Through this we collect direct feedback from the users of the ConE system.

\begin{figure}
\includegraphics[width=0.5\hsize]{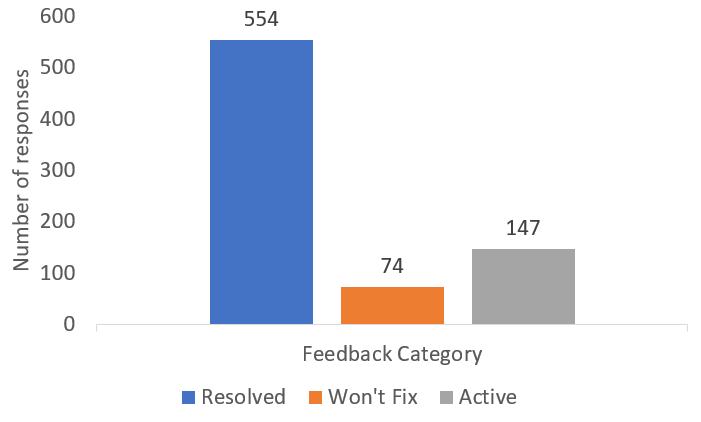}
\vspace{0.5\baselineskip}
\caption{Number of positive (Resolved), negative (Won't Fix), and neutral (Active) responses to \cone notifications}
\Description{554 Resolved, 74 Won't Fix, and 147 Active notifications}
\label{fig:CommentDitribution}
\end{figure}

Figure \ref{fig:CommentDitribution} shows the distribution of the feedback received.
The vast majority (554 out of \NumDecorations, for \PositiveResponseRate\%) of notifications was flagged as ``Resolved''.
For 147 (18.96\%) of the notifications, no feedback was provided.
Various studies have shown that users tend to provide explicit negative feedback when they do not like or agree with a recommendation, while tend not be so explicit about positive feedback \cite{10.1145/2043932.2043957, 10.1145/3289600.3291003}.
Therefore, we cautiously interpret this as neutral to positive.

We manually analyzed all 74 (9.5\%) cases where the developers provided negative feedback.
For the majority of them, the developer was already aware of the other conflicting \pr{}.
In some cases the developers thought that ConE is raising a false alarm as they expect no one else to be making changes to the same files as the ones they are editing. When we show them other overlapping \pr{}s that were active while they were working on their \pr{}, to their surprise, the notification were not false alarms. We list some of the anecdotes in subsection \ref{sec:quotes}.

\begin{figure}
\includegraphics[width=0.75\columnwidth]{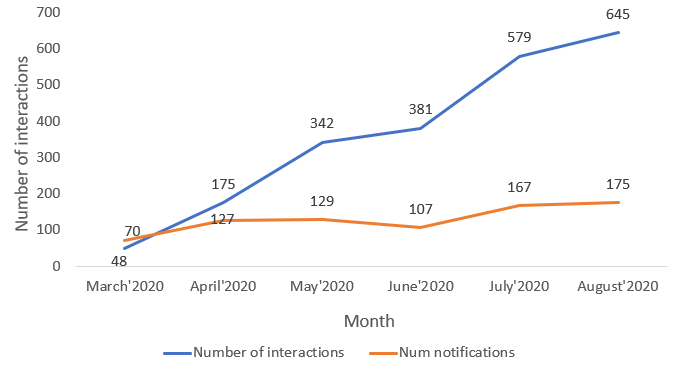}
\caption{Number of notifications ({\color{orange}{orange}}), and number of interactions ({\color{blue}{blue}}) with those notfications, per month. As developers become more familiar with \cone, they increasingly interact with its notifications}
\Description{Notifications per month for six months, fairly stable at 150 per month, and interactions, which grow from 50 to 650 per month}
\label{fig:ClickCount}
\end{figure}

\subsection{Extent of interaction}
As discussed in Section~\ref{Notifications}, a typical ConE notification/comment has multiple elements that a developer can interact with: For each conflicting pull request, the pull request id, files with conflicting changes, and the author name are shown.
These are deep links.
Developers can just take a look at the comment and ignore it or interact with it by clicking on one of the ``clickable elements'' in the ConE notification.
If the user decides to pursue further clicking on one of these elements, that action is also logged as telemetry (in Azure AppInsights).

From March 2020 to September 2020, we logged 2170 interactions on \NumDecorations\ comments that ConE has placed, which amounts to 2.8 clicks per notification on average.
Measured over time, as shown in Figure \ref{fig:ClickCount}, the number of interactions and the ``clicks per notification'' are clearly increasing as more and more people are getting used to ConE comments, and are using it to learn more about conflicting \pr{}s recommended by ConE.

Note that the extent of interaction does not include additional actions developers can take to contact authors of conflicting pull requests once \cone has made them aware of the conflicts, such as reaching out by phone, walking into each other's office, or a simple talk at the water cooler.

\subsection{User Interviews} \label{QualitativeAnalysis}

The quantitative feedback discussed so far captures both direct (comment resolution percentage) and indirect (extent of interaction) feedback.
To better understand the usefulness we directly reached out (via Microsoft Teams, asynchronously) to authors of 100 randomly selected \pr{}s for which ConE placed comments.
The user feedback for these 100 \pr{}s is 45\% positively resolved, 35\% won't fix, and 20\% no response.
The interviewers did not know these authors, nor had worked with them before, also because the teams working on the systems under study are organizationally far away from the interviewers.

The interview format is semi-structured where users are free to bring up their own ideas and free to express their opinions about the ConE system. We posed the following questions:
\begin{enumerate}
    \item Is it useful to know about these other PRs that change the same file as yours?
    \item If yes, roughly how much effort do you estimate was saved as a result of finding out about the overlapping PRs?  If not, is there other information about overlapping PRs that could be useful to you?
    \item Does knowing about the overlapping PRs help you to avoid or mitigate a future merge conflict?
    \item What action (if any) will you likely take now that you know about the overlapping PRs?
    \item Would you be interested in keeping using ConE which notifies you about overlapping PRs in the future? (Note that we aim to avoid being too noisy by not alerting if the overlapping files are frequently edited by many people, if they are not source code files, etc.)
\end{enumerate}

\begin{table}
\setlength\belowcaptionskip{4pt}
\caption{Distribution of qualitative feedback responses}
 \begin{tabular}{c|c|c} 
 \toprule
 & Category & \# of responses \\ \hline
\multirow{2}{*}{Favorable} & I'd love to use ConE & 25 (52.08\%) \\\cline{2-3}
& I will use ConE & 20 (41.67\%) \\ \hline
Unfavorable & I don't want to use ConE & 3 (6.25\%) \\ \hline
\end{tabular}
\label{Tab:Quantative-Feedback}
\end{table}

We did not receive the responses in a uniform format directly based on the structure of the questions.
We used Microsoft Teams to reach out to the developers and the questions are open ended. Therefore, we could not enforce a strict policy on the number of questions the respondents should answer and on the length of the answers. Some of the participants answered all questions, while some answered only one or two.
Some respondents were detailed in their response, while some were succinct with `yes' or `'no' answers.
Some of the respondents provided a free-form response, with an average word count of just 47 words per response. So, we could not calculate the distribution of responses for all questions. However, we see that for question-5, there were responses. We coded and categorized the responses we received for question-5 as explained below.

The first three authors, together, grouped the responses that we received (48 out of 100), until consensus was reached, into two categories: Favorable (if the users would like to continue using ConE, i.e., the answer to question 5 is along the lines of `I will use ConE' or a `I'd love to use/keep using ConE') and Unfavorable (users do not find the ConE system to be useful and do not want to continue using it.). Table \ref{Tab:Quantative-Feedback} shows the distribution of the feedback: 93.75\% of the respondents indicated their interest and willingness to use ConE.

\subsection{Representative Quotes}
\label{sec:quotes}

To offer an impression, we list some typical quotes (positive and negative) that we received from the developers. In one of the \pr{}s where we sent a ConE notification notifying about potential conflicting changes, a developer said: 

\begin{displayquote}
\textit{"I wasn't aware about other 2 conflicting PRs that are notified by ConE. I believe that would be very helpful to have a tool that could provide information about existence of other PRs and let you know if they perform duplicate work or conflicting change!!"}
\end{displayquote}
It turned out that the other two developers (the authors of the conflicting \pr{}s flagged by ConE) are from entirely different organizations and geographies. Their common denominator is the CEO of the company.
It would be very difficult for the author of the reference \pr{} to know about the existence of the other two \pr{}s without ConE bringing it to their notice. 

Several remarks are clear indicators of the usefulness of the ConE system:

\begin{displayquote}
\textit{"Yes, I would be really interested in a tool that would notify overlapping PRs."}
\end{displayquote}

\begin{displayquote}
\textit{"Looking forward to use it! Very promising!"}
\end{displayquote}

\begin{displayquote}
\textit{"ConE is such a neat tool! Very simple but super effective!"}
\end{displayquote}

\begin{displayquote}
\textit{"ConE is a great tool, looking forward to seeing more recommendations from ConE"}
\end{displayquote}

\begin{displayquote}
\textit{"This is an awesome tool, Thank you so much  for working to improve our engineering!"}
\end{displayquote}

\begin{displayquote}
\textit{"It is a nice feature and when altering files that are critical or very complex, it is great to know."}
\end{displayquote}

Some developers mentioned that ConE helped them saving time and/or effort significantly by providing early intervention:
\begin{displayquote}
\textit{"ConE is very useful. It saved at least two hours to resolve the conflicts and smoke again"}
\end{displayquote}

\begin{displayquote}
\textit{"This would save a couple of hours of dev investigation time a month"}
\end{displayquote}

\begin{displayquote}
\textit{"ConE would have saved probably an hour or so for PR <XYZ>"}
\end{displayquote}

We also received feedback from some developers who expressed a feeling that a tool like ConE may not necessarily be useful for their scenarios:
\begin{displayquote}
\textit{"For me no, I generally have context on all other ongoing PRs and work that might cause merge issues. No, thank you!"}
\end{displayquote}

\begin{displayquote}
\textit{"For my team and the repositories that I work in, I don't think the benefit would be that great.
I can see where it could be useful in some cases though"}
\end{displayquote}

\begin{displayquote}
\textit{"It's not helpful for my specific change, but don't let that discourage you. I can see how something like ConE be definitely useful for repositories like <XYZ> which has a lot of common code"}
\end{displayquote}

Another interesting case we noticed is, ConE's ability to help in detecting duplication of work. ConE notified a developer (D1) about an active \pr{} authored by another developer (D2).
After the ConE notification was sent to D1, they realized that D2's \pr{} is already solving the same problem and D2 made more progress. D1 ended up abandoning their \pr{} and pushed several code changes in D2's \pr{}, which was eventually completed and merged. When we reached out to D1, they said:

\begin{displayquote}
\textit{"Due to poor communication / project planning D2 and I ended up working on the same work item. Even if I was not notified about this situation, I would have eventually learned about it, but that would have costed me so much time. This is great!"}
\end{displayquote}

Though we do not observe scenarios like this frequently, this case demonstrates an example of the kind of potential conflicts ConE can surface, in addition to flagging syntactic conflicts.

\begin{table}
\caption{Distribution of quantitative feedback based on size of the repository}

\begin{tabular}{l|r|r|r}
\toprule
Feedback             & Large repositories & Small repositories & Total \\
\midrule
Positively resolved  & 404 (77.69\%)      & 150 (58.82\%)      & 554 (71.48\%)  \\
Won't fix            &  33 (\ \ 6.34\%)   &  41 (16.08\%)      &  74 (\ \ 9.54\%) \\
No response          &  83 (15.96\%)      &  64 (25.10\%)      & 147 (18.96\%)  \\
\midrule
Total                & 520 (67.09\%)      & 255 (32.90\%)      & 775 (100.0\%)  \\
\bottomrule
\end{tabular}
\label{Tab:Quantative-Feedback-Distribution}
\end{table}

\subsection{Factors Affecting \cone Appreciation}
After analyzing all the responses from our interviews, analyzing the \pr{}s on which we received `Won't Fix' and interviewing respective \pr{} authors, we identified the following main factors as to what makes a developer incline towards using a system like ConE.

\paragraph{Developers who found the ConE notifications useful:}
These are the developers who typically work on large services with distributed development teams across multiple organizations, geographies and time zones. They also tend to work on core platforms or common infrastructure (as opposed to the ones who make changes to the specific components of the product or service).
To corroborate this, the first author classified the repositories into large and small manually, based on the size and the activity volume in those repositories.
We then, programmatically, categorized the 628 responses based on their repository sizes.
The results, in Table~\ref{Tab:Quantative-Feedback-Distribution}), show that for large repositories developers are positive for 77.69\% (404/520) of the cases, whereas for small repositories this is 58.82\% (150/255).

\paragraph{Developers who found ConE not so useful:}
These developers are the ones who work on small micro services or small scale products and typically work in smaller teams. These developers, and their teams, tend to have delineated responsibilities.
They usually have more control over who makes changes to their code base.
Interestingly, there were cases where some of these developers were surprised to see another active \pr{}, created by a different developer, from a different team sometimes, which was editing the same area of the source code as their \pr{}. This could be a result of underestimating the pace with which service dependencies are introduced, product road maps change, and codebases are re-purposed in large scale organizations.

\section{Discussion} \label{Scope}

In this section we describe the outlook and future work. We also explain some of the limitations of the ConE system and how we plan to address them. 

\subsection{Outlook} One of the immediate goals of the ConE system is to expand its reach beyond the initial \NumRepos\ on which it is enabled, and eventually on every source code repository in Microsoft. Furthermore, in the long run, Microsoft may consider offering ConE as part of its Azure DevOps pipeline, making it available to its customers across the world. Likewise, GitHub may consider to develop a free version of ConE as an extension on the GitHub marketplace for the broader developer community to benefit from this work.

As explained, ConE is expected to generate false alarms because of the fact that it is a heuristics based system.To improve the system and reduce the number of false alarms at this point, ConE checks for very simple but effective heuristics (see Section~\ref{ConEAlgo}) and conditions to flag conflicting changes that causes unintended consequences. We offer three configuration parameters (see Section~\ref{knobs}), that help us make the solution effective by striking a suitable balance between the rate of false alarms and coverage, and customize the solution based on individual repository needs.

To further improve precision, we would like to investigate the options that let us go one level deeper from file level to, e.g., analyze actual code diffs. Understanding code diffs and performing semantic analysis on them is a natural next a step for a system like ConE. 
Providing diff information across every developer branch is fundamentally expensive, so it is not offered by Azure DevOps, the source control system on which ConE is operationalized, nor by  other commercial or free source control systems like GitLab or GitHub.
A possible remedy is to bring the diff information into the ConE system. This involves checking out two versions of the same file, within \cone, and finding differences. This has to happen in real-time, in a scalable and language agnostic fashion. 

Once we have the diff information, another idea is to apply deep learning and code embeddings to develop better contextual understanding of code changes. We can use the semantic understanding in combination with the historical data about concurrent and non-concurrent edits to develop better prediction models and raise alarms when concurrent edits are problematic.

ConE was found to be useful by facilitating early intervention about the potential conflicting change. However, this does not fully solve the problem i.e., fixing the merge conflicts or merging the duplicate code. Exploring auto-fixing of conflicts or code duplication as a natural extension to ConE's conflict detection algorithm will help in alleviating the problems caused by the conflicts and fixing them in an automated fashion.

\subsection{Threats to validity}
Concerning internal validity, our qualitative analysis was conducted by reaching out to the developers via Microsoft Teams, asynchronously. None of the interviewers know the people that were reached out neither worked with them before. We purposefully avoided deploying ConE on repositories that are under the same organization as any of the researchers involved in this work. As Microsoft is a huge company and most of the users of the ConE service are organizationally distant from the interviewers, the risk of response bias is very minimal. However, there is a small chance that respondents may be positive about the system because they want to make the interviewers, who are from the same company, happy.

Concerning external validity, the empirical analysis, design and deployment, evaluation and feedback collection are done specifically in the context of Microsoft. The correlations we reported in Table \ref{Tab:Correlation-Comparison} can vary based on the setting of the organization in which the analysis is performed. As Microsoft is one of the world's largest concentration of developers, and developers at Microsoft uses very diverse set of tools, frameworks, programming languages, our research and the ConE system will have a broader applicability. However, at this point the results are not verified in the context of other organizations.

\section{Conclusion and future work}

In this paper, we seek to address problems originating from concurrent edits to overlapping files in different pull requests.
We start out by exploring the extent of the problem, establishing a statistical relationship between concurrent edits and the need for bug fixes in six active industrial repositories from Microsoft.

Inspired by these findings we set out to design \cone, an approach to detect concurrently edited files in pull requests at scale.
It is based on heuristics like the extent of overlap and the presence of rarely concurrently edited files between pairs of pull requests.
To make sure the precision of the system is sufficiently high, we deploy
various filters and parameters that help in controlling the behavior of the ConE system.

ConE has been deployed on \NumRepos\ repositories inside Microsoft.
During a period of six months, \cone generated \NumDecorations notifications, from which \PositiveResponseRate\% received positive feedback.
Interviews with 48 developers showed 93\% favorable feedback, and applicability in avoiding merge conflicts as well as duplicate work.

In the future, we anticipate \cone will be employed at substantially more systems within Microsoft.
As ConE has been deployed and found to be useful by the developers in a large and diverse (in terms of programming languages used, tools, engineering systems, geographical presence, etc) organization like Microsoft, we believe the techniques and the system has applicability beyond Microsoft.
Furthermore, we see opportunities for implementing a \cone service for systems like GitHub or GitLab.
Future research surrounding \cone might entail improving its precision by learning from past user feedback or by leveraring diffs without sacrificing scalability.
Beyond warnings, future research could also target automating actions to be taken to address the pull request conflicts detected by \cone.

\bibliographystyle{ACM-Reference-Format}
\bibliography{Paper}

\end{document}